\documentclass{article}
\usepackage{spconf,amsmath,graphicx,bm,amssymb}
\usepackage{algorithm}
\usepackage{color}

\usepackage{graphicx,subfigure}

\usepackage[pdfstartview=FitH,
CJKbookmarks=true,
bookmarksnumbered=true,
bookmarksopen=true,
colorlinks,
pdfborder=001,
linkcolor=blue,
anchorcolor=blue,
citecolor=blue,
]{hyperref}
\hypersetup{hidelinks}

\usepackage{algorithmic}

\newcommand{\ba}{\mathbf{a}}

\newcommand{\bh}{\mathbf{h}}

\newcommand{\bn}{\mathbf{n}}

\newcommand{\bx}{\mathbf{x}}

\newcommand{\bs}{\mathbf{s}}

\newcommand{\by}{\mathbf{y}}

\newcommand{\bA}{\mathbf{A}}

\newcommand{\bH}{\mathbf{H}}

\newcommand{\bP}{\mathbf{P}}

\newcommand{\RR}{\mathcal{R}}

\newcommand{\T}{{{\mathsf{T}}}}

\newcommand{\I}{\mathcal {I}}

\newtheorem{dingli}{Theorem~}
\newtheorem{yinli}{Lemma~}

\def\x{{\mathbf x}}

\title{Efficient Quantized Constant Envelope Precoding for Multiuser Downlink Massive MIMO Systems}
\name{Zheyu Wu$^{\S,\star}$,  Ya-Feng Liu$^{\S}$, Bo Jiang$^{\dag}$, and Yu-Hong Dai$^{\S}$
}

\address{$^{\S}$LSEC, ICMSEC, AMSS, Chinese Academy of Sciences, Beijing, China \\[2pt]
$^{\star}$School of Mathematical Sciences, University of Chinese Academy of Sciences, Beijing, China\\[2pt]
    $^{\dag}$School of Mathematical Sciences, Nanjing Normal University, Nanjing, China\\[2pt]
  Email:  \{wuzy, yafliu, dyh\}@lsec.cc.ac.cn, jiangbo@njnu.edu.cn }
  
\begin{document}

\ninept
\maketitle
%




%
%

\begin{abstract}
Quantized constant envelope (QCE) precoding, a new transmission scheme that only discrete QCE transmit signals are allowed at each antenna, has gained growing research interests due to its ability of reducing the hardware cost and the energy consumption of massive multiple-input multiple-output (MIMO) systems.  However, the discrete nature of QCE transmit signals greatly complicates the precoding design. In this paper, we consider the QCE precoding problem for a massive MIMO system with phase shift keying (PSK) modulation and develop an efficient approach for solving the constructive interference (CI)  based problem formulation.   Our approach is based on  a custom-designed (continuous) penalty model that is  equivalent to the original discrete problem. Specifically, the penalty model relaxes the discrete QCE constraint and penalizes it in the objective with a negative $\ell_2$-norm term, which leads to a non-smooth non-convex optimization problem. To tackle it, we resort to our recently proposed alternating optimization (AO) algorithm. We show that the AO algorithm admits closed-form updates at each iteration when applied to our problem and thus can be efficiently implemented. Simulation results demonstrate the superiority of the proposed approach over the existing  algorithms.

\end{abstract}
\vspace{-0.1cm}
\section{Introduction}\vspace{-0.2cm}
High hardware cost and power consumption have been widely recognized as the major issues for the practical deployment of massive multiple-input multiple-output (MIMO) systems \cite{massivemimo2}. To deal with this, various techniques have been proposed, including  employing low-resolution digital-to-analog converters (DACs) at the antenna arrays \cite{SQUID} and reducing the peak-to-average power ratio (PAPR) of the transmit signals (which enables the use of power-efficient power amplifiers (PAs)) \cite{PAbook}. The combination of the above two techniques motivates a new transmission scheme  in the framework of symbol-level precoding \cite{symboltut,CItutorial} called quantized constant envelope (QCE) transmission, where each antenna is restricted to transmit coarsely QCE signals. 

The QCE precoding problem was first formulated based on the classical minimum mean square error (MMSE) criterion \cite{trellis}--\cite{ofdm}. Recently, it has been realized that incorporating the idea of constructive interference (CI) into precoding design can greatly improve the bit error rate (BER)  performance of the system \cite{CItutorial,CI2}.
Motivated by this, the authors in \cite{ciqce} adopted the CI metric to formulate the QCE precoding problem.  In addition to the MMSE and CI metrics, some works also considered the symbol error probability (SEP) criterion directly (for quadrature amplitude modulation (QAM))  \cite{GEMM,GEMM2}. In this paper, we focus on the CI-based model due to its good BER performance and its computational tractability.  It is worthwhile mentioning a widely studied special case of QCE precoding, one-bit precoding, where the finite phases allowed to be transmitted by each antenna is $\{\frac{(2i-1)\pi}{4}, i=1,2,3,4\}$. 
Although numerous algorithms have been proposed  for one-bit precoding \cite{CImodel}--\cite{onebitalg}, very few algorithms have been designed for solving the general QCE precoding problem.  To the best of our knowledge, the only CI-based algorithm designed in the  QCE context is the maximum safety margin (MSM) algorithm \cite{ciqce}, which directly relaxes the discrete QCE constraint and suffers from poor BER performance. We also remark here that the direct extension of the existing algorithms for one-bit precoding to solve the QCE precoding problem either is difficult (due to the more general and difficult QCE constraint) 
or leads to a poor performance. This remark will become clear in the algorithmic design and simulation result sections.  

 In this paper, we consider the general QCE precoding design problem for a downlink massive MIMO system with phase shift keying (PSK) signaling and propose a new penalty approach to solve the CI-based problem formulation. 
More specifically, we first propose a penalty model exploiting the special structure of the discrete QCE constraint and establish its global equivalence with the original problem. Then, we apply the alternating optimization (AO) algorithm proposed in our recent work \cite{journal} to solve the penalty model. By carefully investigating the update rule, we find that each iteration of the AO algorithm admits closed-form solutions, making it suitable to solve large-scale problems arising from the massive MIMO scenario. Simulation results show that, compared to the existing algorithms, our proposed approach achieves better BER performance with lower computational cost.  In addition, the simulation also indicates that the BER performance of the system can be significantly enhanced by slightly increasing the resolution of DACs from 1 bit to 2-4 bits. 
\vspace{-0.4cm}\section{problem formulation}
\vspace{-0.1cm}
\subsection{System Model}\vspace{-0.1cm}
Consider a massive MIMO downlink transmission scenario, where an $N$-antenna  base station (BS) serves $K$ single-antenna users simultaneously.  Let $\mathbf{t}\in\mathbb{C}^N$ denote the transmitted signal vector from the BS. Then the received signal at the users is given by\vspace{-0.1cm}
$$\vspace{-0.1cm}\by=\bH\mathbf{t}+\bn,$$
where $\bH=[\bh_1,\bh_2,\ldots,\bh_K]^\mathsf{T}\in\mathbb{C}^{K\times N}$ is the channel matrix between the BS and the users; $\bn\sim\mathcal{CN}(0,\sigma^2\mathbf{I})$ models the additive Gaussian noise.  In this paper, we consider QCE transmission, i.e., each element of $\mathbf{t}$ can only be selected from a finite set of symbols with constant amplitude. Let $\mathbf{t}=\sqrt{\frac{P_T}{N}}\bx_T$, where $P_T$ is the total transmit power at the BS and $\x_T$ is the normalized transmitted signal vector. Then the QCE constraint can be expressed as 
\vspace{-0.2cm}$$\x_T(i)\in\mathcal{X}_L\triangleq\left\{e^{j\frac{(2l-1)\pi}{L}},~l=1,\dots, L\right\},~i=1,2,\dots, N,\vspace{-0.2cm}$$ where $j$ is the imaginary unit and 
$L$ is the number of quantization levels.   Note that $L$-level quantization corresponds to $(\log_2 L-1)$-bits quantization of DACs. In particular, when $L=4,$ the QCE constraint reduces to the one-bit constraint.

Let $\bs=[s_1,s_2,\dots, s_K]^\mathsf{T}$ be the intended symbol vector for the users, where each $s_i$ is assumed to be drawn from the $M$-PSK constellation set. The QCE precoding problem considered in this paper is to design the transmitted signal vector $\x_T$ (based on $\bs$ and $\bH$) such that $\bs$ can be recovered by the users.
\vspace{-0.3cm}
\subsection{CI-Based Problem Formulation}
In this paper, we adopt the CI metric to formulate our problem. Roughly speaking, the CI metric aims to maximize the distance between the (noiseless) received signal and the decision boundary of the intended symbol. 
A popular CI-based model is the symbol scaling model \cite{CImodel}, which was originally proposed in the context of one-bit precoding  and will be generalized to formulate the  considered QCE precoding problem below.
\begin{figure}
\centering
\includegraphics[scale=0.095]{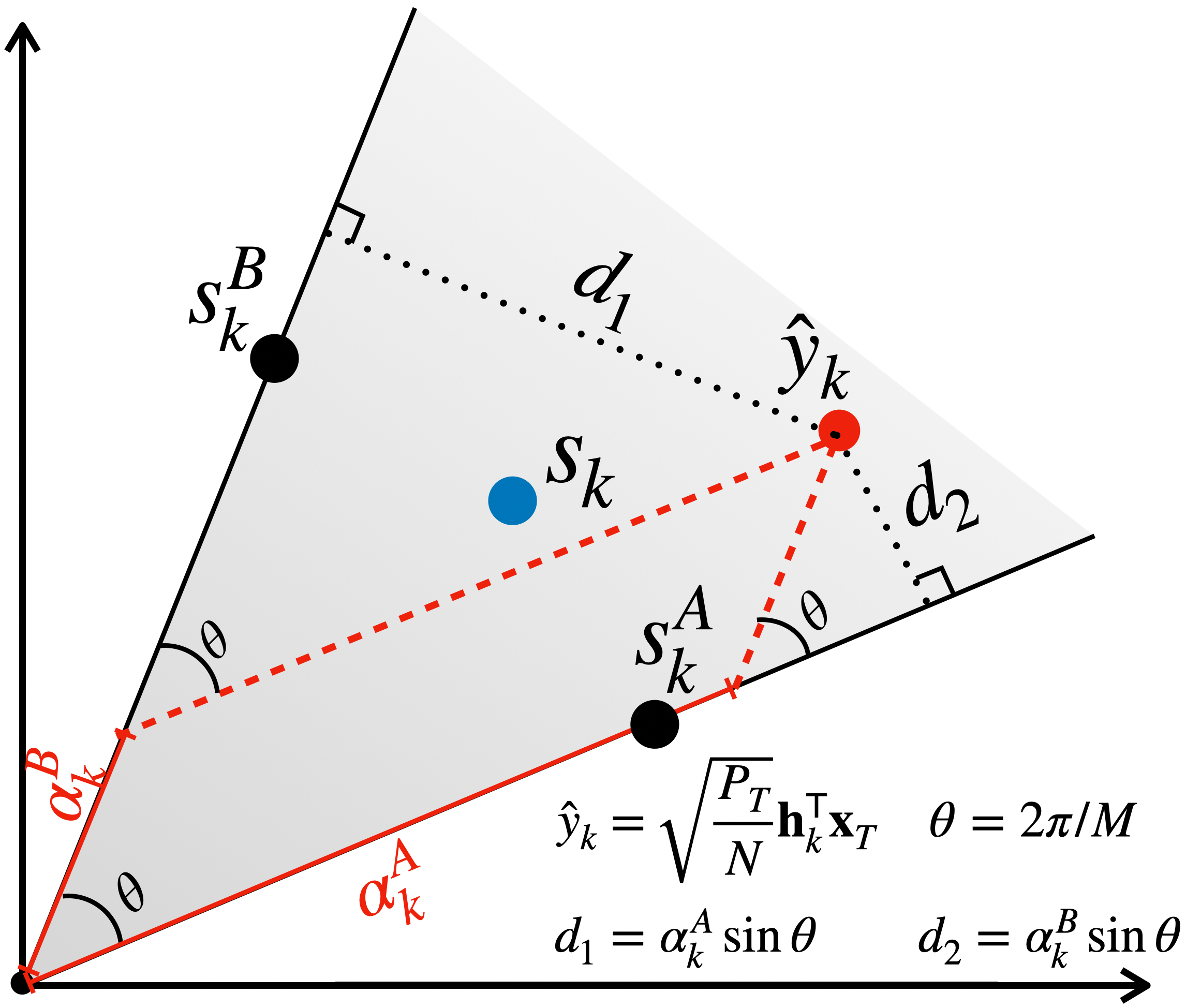}
\vspace{-0.25cm}
\caption{An illustration of the CI formulation for 8-PSK.}\vspace{-0.5cm}
\label{fig3}
\end{figure}

  To illustrate the main idea of the symbol scaling model, we depict in Fig. \ref{fig3} a piece of decision region for 8-PSK modulation, where $s_k$ is the intended symbol and $\hat{y}_k=\sqrt{\frac{P_T}{N}}\bh_k^\mathsf{T}\x_T$ is the noiseless received signal. We decompose $\hat{y}_k$ along the decision boundaries of $s_k$ as 
$\hat{y}_k=\alpha_k^As_k^A+\alpha_k^Bs_k^B,$
where $s_k^A=s_k e^{-j\frac{\pi}{M}}$ and $s_k^B=s_ke^{j\frac{\pi}{M}}$ are the unit vectors on the decision boundaries. With this decomposition, the distance between $\hat{y}_k$ and the decision boundaries of $s_k$ can be expressed as\vspace{-0.2cm}   $$\min\{d_1,d_2\}=\min\{\alpha_k^A,\alpha_k^B\}\sin\frac{2\pi}{M},\vspace{-0.15cm}$$
i.e., the distance is proportional to $\min\{\alpha_k^A,\alpha_k^B\}$. This motivates the following symbol scaling model \cite{CImodel}:\vspace{-0.15cm}
\begin{equation*}
\begin{aligned}
\max_{\bx_T}~&\min_{k\in\{1,2,\dots,K\}}~\left\{\alpha_k^A,\alpha_k^B\right\}\\
\hspace{-0.3cm}\text{(}\text{P}_0\text{)}\hspace{0.3cm}\text{s.t. }~&\sqrt{\frac{P_T}{N}}\bh_k^\T\bx_T=\alpha_k^As_k^A+\alpha_k^Bs_k^B,\quad k=1,2,\dots,K,\\
~&\bx_T(i)\in\mathcal{X}_L, \quad i=1,2,\dots, N,
\end{aligned}
\end{equation*} 
where the objective is to maximize the minimum distance between the (noiseless) received signal and the corresponding decision boundary among all users. With similar manipulations as in \cite{CImodel} {\color{black}(by rewriting (P$_0$) into the real space, rearranging the variables, and transforming maximization into minimization)}, the above problem can be equivalently expressed in a more compact form as \vspace{-0.1cm}
\begin{equation*}\label{Pe}
\text{(P)}\quad
\begin{aligned}
\min_{\x}~&\max_{m\in\{1,2,\dots,2K\}}\mathbf{a}_m^\T\x\\
\text{s.t. }~&\x_i\in \mathcal{X}_L^\RR\triangleq\left\{\left[\begin{matrix}\cos\frac{(2l-1)\pi}{L}\\\sin\frac{(2l-1)\pi}{L}\end{matrix}\right],~ l=1,\dots, L\right\},~i=1,\dots, N,
\end{aligned}
\end{equation*}
where $\x$ is decomposed as $\x=[\bx_1,\dots,\bx_N]^\mathsf{T}$ with each $\bx_i=[x_{2i-1},x_{2i}]^\mathsf{T}$;
$\{\mathbf{a}_m\}_{1\leq m\leq 2K}$ is a sequence of problem-dependent vectors whose explicit expressions are given by
$$
\begin{aligned}
\left[\begin{matrix}\ba_{2k-1}^\mathsf{T}\\\ba_{2k}^\mathsf{T}\end{matrix}\right]=\frac{-1}{\sin\frac{2\pi}{M}}\sqrt{\frac{P_T}{N}}\left[\begin{matrix}\I(s_k^B)&-\RR(s_k^B)\\-\I(s_k^A)&\RR(s_k^A)\end{matrix}\right]\left[\begin{matrix}h_{k,1}^\RR,\dots,h_{k,N}^\RR\end{matrix}\right],
\end{aligned}$$
where $h_{k,i}^\RR=\left[\begin{smallmatrix}\RR(h_{k,i})&-\I(h_{k,i})\\\I(h_{k,i})&\RR(h_{k,i})\end{smallmatrix}\right]$, $\RR(\cdot)$ and $\I(\cdot)$ refer to the real and imaginary parts of their arguments, respectively. 
\section{Proposed Approach for Solving (P)}
Problem (P) is a large-scale non-smooth optimization problem with complicated discrete constraints. In this section, we propose an efficient algorithm for solving (P) by exploiting its special structure. 

\vspace{-0.1cm}
\subsection{Penalty Approach for Solving Problem (P)} 
\begin{figure}[t]\vspace{-0.2cm}
\includegraphics[scale=0.09]{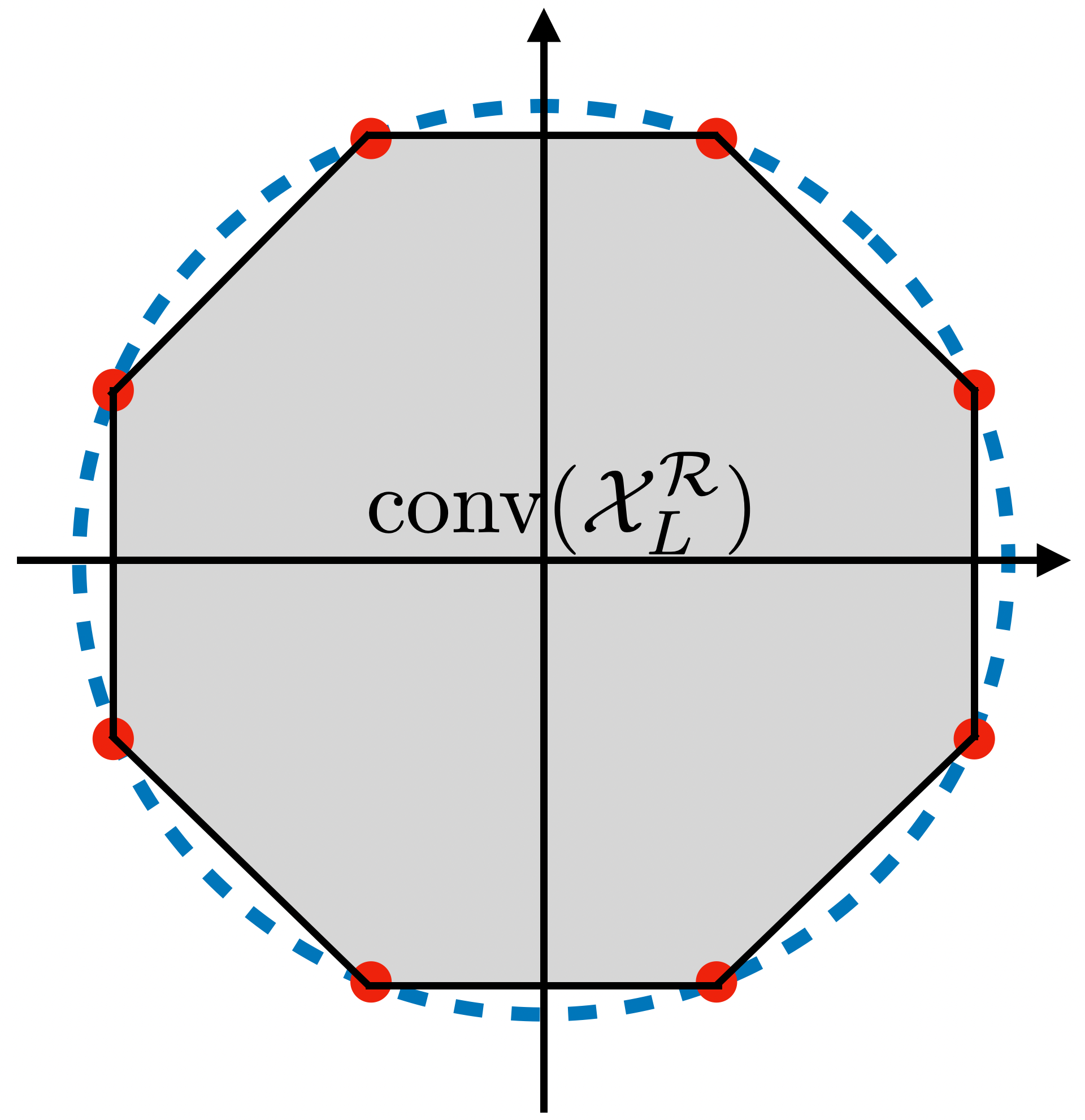}
\centering
\vspace{-0.4cm}
\caption{An illustration of $\text{conv}(\mathcal{X}_L^\RR)$ for $L=8$.}\vspace{-0.4cm}
\label{convexhull}
\end{figure}

A straightforward way of dealing with the discrete constraints  $\x_i\in \mathcal{X}_L^\RR$ in (P) is to relax them into their convex hulls $\x_i\in \text{conv}(\mathcal{X}_L^\RR)$, which transforms the discrete problem  into a continuous one. However, this will greatly expand the feasible region of the original problem and in most cases solving the relaxation problem gives a solution lying inside the convex hull, i.e., not satisfying the discrete constraints. To ensure a feasible solution of (P) while still obtaining a continuous model, we resort to the penalty technique and consider the following penalty model:\vspace{-0.4cm}
\begin{equation*}
\text{(}\text{P}_\lambda\text{)} \qquad
\begin{aligned}
\min_{\x}~&\max_{m\in\{1,2,\dots,2K\}}~\mathbf{a}_m^\T\x-\lambda\sum_{i=1}^N\|\x_i\|_2\\
\text{s.t. }~&\x_i\in\text{conv}(\mathcal{X}_L^\RR),~i=1,2,\dots, N.
\end{aligned}
\end{equation*}
The (negative) penalty term in (P$_\lambda$) is  introduced based on the observation that feasible points of (P) are those with the largest $\ell_2$-norm in $\text{conv}(\mathcal{X}_L^\RR)$; see Fig. \ref{convexhull}. Hence, by encouraging large value of $\|\x_i\|_2$, the penalty term forces each $\x_i$ to be feasible for (P). Similar ideas have also been considered in \cite{GEMM, conference, journal}. In particular, our recent works \cite{conference,journal} considered the one-bit precoding problem and proposed to add a negative $\ell_1$ norm in the objective function to promote one-bit solutions. The penalty model (P$_\lambda$) in this paper can be regarded as a generalization of the negative $\ell_1$ penalty model in \cite{conference,journal}. It is worthwhile highlighting here that, unlike the one-bit case, the negative $\ell_1$ penalty model in \cite{conference,journal} is generally not (globally) equivalent to the original problem in the QCE case.
This is a key difference between one-bit and QCE precoding problems in algorithmic design and is the main reason why the $\ell_2$ norm is used in (P$_\lambda$). 

In the following theorem, we establish the relationship between global and local solutions of problems (P) and (P$_\lambda$).

\begin{dingli}\label{theorem3}
If the penalty parameter $\lambda$ in (P$_{\lambda}$) satisfies $\lambda>\frac{\sin\frac{\pi}{L}}{1-\cos\frac{\pi}{L}}\max_m\|\mathbf{a}_m\|_2$, then the following results hold:
\begin{enumerate}
\renewcommand{\labelenumi}{(\theenumi)}
\item Problems (P) and (P$_\lambda$) have the same optimal solutions;
\item Every local minimizer of (P$_\lambda$) lies on the boundary of $\text{conv}(\mathcal{X}_L^\RR)$.
\end{enumerate}
\end{dingli} \vspace{-0.2cm}
The first result in Theorem \ref{theorem3} establishes the global equivalence between (P$_\lambda$) and (P), 
which shows that the transformation from (P) to (P$_\lambda$) makes sense. However, it is generally hard to obtain the global solution of a non-smooth non-convex problem like (P$_\lambda$). For a well designed iterative algorithm applied to solve problem (P$_\lambda$), the best that one can expect is that the algorithm can find a local minimizer of (P$_\lambda$). 
Unfortunately, the second result in Theorem \ref{theorem3} shows that even if a local minimizer is obtained, it might not be a feasible solution of (P), which is in sharp contrast to the one-bit case \cite[Theorem 3]{journal} and poses further challenge on algorithmic design. In the next subsection, we will apply our recently proposed AO algorithm to solve (P$_\lambda$), and by carefully investigating the update rule and designing the stopping criterion, the algorithm is guaranteed to terminate at a feasible solution of (P). 

We emphasize here that transforming (P) into (P$_\lambda$) not only benefits the algorithmic design, but also helps to obtain a high-quality solution. More specifically, by gradually increasing the penalty parameter $\lambda$ and tracing the solution path, we can bypass bad local minimizers of (P$_\lambda$) (i.e., bad solutions of (P)). Such homotopy technique is widely applied in literature \cite{GEMM,conference,journal}.  We summarize the proposed penalty approach combined with the homotopy technique for solving (P) as follows.
 \vspace{-0.2cm}
\begin{algorithm}[H]
	\caption{Proposed penalty approach for solving problem (P)}
	\begin{algorithmic}[1]
		\STATE \textbf{Initialize:} $\lambda^{(0)}$, $\delta>1$, $\bx^{(0)}$, and $t=0.$
		\REPEAT
		\STATE  Apply the AO algorithm (in the next subsection) to solve the penalty problem (P$_{\lambda^{(t)}}$) with initial point $\bx^{(t)}$ and let the output be $\bx^{(t+1)}$; set $\lambda^{(t+1)}=\delta\lambda^{(t)}$.
		\STATE 
		Set $t \leftarrow t+1.$
		\UNTIL{$\bx^{(t)}$ is a feasible solution of (P)}.
		\STATE \textbf{Output:}  $\bx^{(t)}$.
	\end{algorithmic} \label{alg:framework}
\end{algorithm}\vspace{-0.7cm}
\subsection{AO Algorithm for Solving  (P$_\lambda$)}\vspace{-0.1cm}
In this subsection, we apply the AO algorithm proposed in \cite{journal} to solve the penalty problem (P$_\lambda$). To begin, we  transform (P$_\lambda$) into the following min-max problem:\vspace{-0.4cm}
\begin{equation*}\vspace{-0.2cm}
\text{(}\widehat{\text{P}}_\lambda\text{)}\quad\min_{\bx\in(\text{conv}(\mathcal{X}_L^{\RR}))^N}~{\max_{\by\in\Delta}~\by^\T\bA\bx}-\lambda\sum_{i=1}^N\|\bx_i\|_2,
\end{equation*}
where $\bA=[\mathbf{a}_1,\dots,\mathbf{a}_{2K}]^\mathsf{T}$ and $\Delta=\{\by\mid\mathbf{1}^\mathsf{T}\by=1, \by\geq \mathbf{0}\}$.
Given $\bx^{(k)}$ and $\by^{(k)}$, the AO algorithm \cite{journal} alternately updates the variables $\bx$ and $\by$ at the $(k+1)$-th iteration as follows:
 \vspace{-0.3cm}
\begin{subequations}
\begin{align}\vspace{-0.2cm}
\hspace{-0.1cm}\bx^{(k+1)}\hspace{-0.1cm}&\in\arg\hspace{-0.4cm}\min_{\bx\in(\text{conv}(\mathcal{X}_L^\RR))^N}\hspace{-0.1cm}{\by^{(k)}}^\T\hspace{-0.05cm}\bA\bx\hspace{-0.05cm}-\hspace{-0.05cm}\lambda\hspace{-0.05cm}\sum_{i=1}^N\|\bx_i\|_2\hspace{-0.05cm}+\hspace{-0.05cm}\frac{\tau_k}{2}\|\bx-\bx^{(k)}\|_2^2,\label{updatex}\\
\by^{(k+1)}\hspace{-0.1cm}&=\text{Proj}_{\Delta}\left(\by^{(k)}+{\rho_k}\bA\bx^{(k+1)}-{\rho_k}c_k\by^{(k)}\right),\label{updatey}
\end{align}
\end{subequations}
where $\{\tau_k\}, \{\rho_k\}, \{c_k\}$ are the algorithm parameters.
The convergence of the AO algorithm has been established in \cite{journal}.  

Next, we will focus on the implementation details of \eqref{updatex} and \eqref{updatey}. 
The $\by$-update in  \eqref{updatey}  is exactly the same as that in \cite{journal}, which can be implemented very efficiently. The key difference lies in the update of $\bx$ in  \eqref{updatex}. Note that in \cite{journal}, the one-bit constraint and the negative $\ell_1$ penalty are both fully separable in all components of $\bx$, and thus the corresponding $\bx$-subproblem can be decomposed into a group of  one-dimensional problems. 
Unlike \cite{journal}, the $\bx$-subproblem in \eqref{updatex} can only be decomposed into a group of two-dimensional problems with variable $\bx_i=[x_{2i-1},x_{2i}]^\mathsf{T},$ which cannot be further decomposed into two one-dimensional problems because the two variables  $x_{2i-1}$ and $x_{2i}$  are coupled together in both the constraint and the penalty term in (P$_\lambda$).   How to efficiently solve the corresponding two-dimensional problems is the main challenge for the efficient implementation of the AO algorithm. Fortunately, we can verify that those two-dimensional problems admit closed-form solutions.

\begin{figure}[t]
\centering    \vspace{-0.8cm}
\includegraphics[scale=0.083]{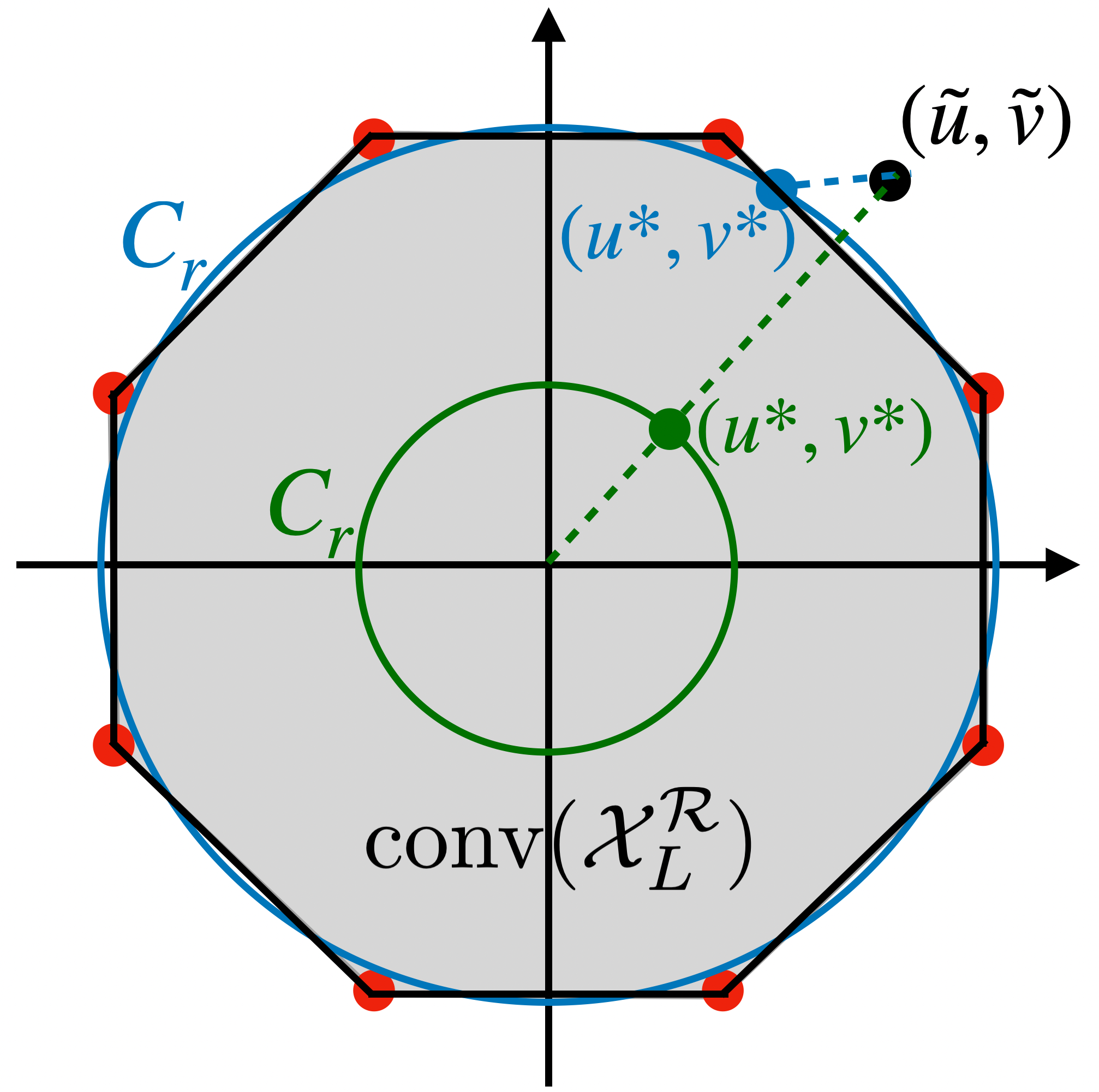}      
\vspace{-0.3cm}
\caption{An illustration of the solution structure of the inner minimization problem over $[u,v]$ in \eqref{eqn:sol2}.}\vspace{-0.55cm}
\label{sol_fig2}
\end{figure}

To be more specific, we note that \eqref{updatex} can be decomposed into  \hspace{-0.05cm} $N$ \hspace{-0.05cm} problems of the following form (with a positive scaling factor being ignored): \vspace{-0.2cm}
\begin{equation}\label{eqn:sol1}
[u^*,v^*]\in\arg\hspace{-0.1cm}\min_{[u,v]\in\text{conv}(\mathcal{X}_L^\RR)}\hspace{-0.05cm}(u-\tilde{u})^2\hspace{-0.04cm}+(v-\tilde{v})^2-\beta \sqrt{{u}^2+{v}^2}.
\vspace{-0.2cm}
\end{equation} 
Introducing an intermediate variable $r$, the optimization problem in \eqref{eqn:sol1} can be equivalently expressed as\vspace{-0.1cm}
\begin{equation}\label{eqn:sol2}
\min_{0\leq r\leq 1}\min_{u^2+v^2=r^2\atop [u,v]\in\text{conv}(\mathcal{X}_L^\RR)}~(u-\tilde{u})^2+(v-\tilde{v})^2-\beta r.
\vspace{-0.16cm}
\end{equation} 
From a geometric point of view,  the inner minimization problem over $(u,v)$ in problem \eqref{eqn:sol2} (with given $r$)
 is to project $[\tilde{u},\tilde{v}]$ onto  set $\text{conv}(\mathcal{X}_L^\RR)\cap C_r$, where $C_r=\{[u,v]\mid u^2+v^2=r^2\}$.  
Let $[{u}^{(r)},{v}^{(r)}]$ be the projection of $[\tilde{u},\tilde{v}]$  onto $C_r$. Then, 
if $r$ is small enough such that  $[{u}^{(r)},{v}^{(r)}]$ lies strictly inside $\text{conv}(\mathcal{X}_L^\RR)$, {\color{black}the optimal solution of the inner problem in \eqref{eqn:sol2} is exactly} $[u^{(r)},v^{(r)}]$;
  otherwise, $C_r \cap \text{bnd}(\text{conv}(\mathcal{X}_L^\RR))$ is a nonempty finite set and {\color{black}the optimal solution} is the point in $C_r \cap \text{bnd}(\text{conv}(\mathcal{X}_L^\RR))$ that is closest to $[\tilde{u}, \tilde{v}],$ where $  \text{bnd}(\text{conv}(\mathcal{X}_L^\RR))$ denotes the boundary of  $\text{conv}(\mathcal{X}_L^\RR)$;
 see Fig. \ref{sol_fig2} for an illustration of the solution structure of the inner minimization problem over $[u,v]$ in \eqref{eqn:sol2}. 

{\color{black} Now we are ready to give the following result}, which characterizes the structure of the optimal solution to problem \eqref{eqn:sol1}. \vspace{-0.13cm}
\begin{yinli}\label{lem1}
Let $\left[\begin{matrix}\tilde{u}_{\alpha}\\\tilde{v}_{\alpha}\end{matrix}\right]=\bP_\alpha\left[\begin{matrix}\tilde{u}\\\tilde{v}\end{matrix}\right],$  where 
$\mathbf{P}_{\alpha}=\left[\begin{matrix}\cos\alpha&\sin\alpha\\-\sin\alpha&\cos\alpha\end{matrix}\right]$ and $\alpha=\left\lfloor{\frac{\arg(\tilde{u}+j\tilde{v})+\frac{\pi}{L}}{\frac{2\pi}{L}}}\right\rfloor\frac{2\pi}{L}$, and let $h(r)$ denote the optimal value of the inner minimization problem over $[u,v]$ with given $r$ in \eqref{eqn:sol2}. Then
\begin{equation*}\vspace{-0.2cm}
h(r)\hspace{-0.1cm}=\hspace{-0.1cm}\left\{\hspace{-0.05cm}
\begin{matrix}(r-\sqrt{\tilde{u}_{\alpha}^2+\tilde{v}_{\alpha}^2})^2-\beta r,&\text{if }~0\leq r \leq r_0;\vspace{0.15cm}\\
\left(\cos\frac{\pi}{L}-\tilde{u}_{\alpha}\right)^2\hspace{-0.15cm}+\hspace{-0.05cm}\left(\sqrt{r^2-\cos^2\frac{\pi}{L}}-|\tilde{v}_{\alpha}|\right)^2\hspace{-0.1cm}-\beta r,\hspace{-0.2cm}&\text{if }~r_0< r\leq 1,
\end{matrix}
\right.\vspace{0.1cm}
\end{equation*}
where $r_0=\frac{\sqrt{\tilde{u}_{\alpha}^2+\tilde{v}_{\alpha}^2}}{\tilde{u}_{\alpha}}\cos \frac{\pi}{L}$, and 
$\left[\begin{matrix}u^*\\ v^*\end{matrix}\right]=\bP_\alpha^{-1}\left[\begin{matrix}{u}_{\alpha}^*\\{v}_{\alpha}^*\end{matrix}\right],$
where
$$\vspace{-0.1cm}\left[\begin{matrix}{u}_{\alpha}^*\\ {v}_{\alpha}^*\end{matrix}\right]=\left\{
\begin{aligned} \frac{r^*}{\sqrt{\tilde{u}_{\alpha}^2+\tilde{v}_{\alpha}^2}}\,\left[\tilde{u}_{\alpha},\tilde{v}_{\alpha}\right]^\mathsf{T},\quad\qquad&\text{if }~0\leq r^*\leq r_0;\\
\left[\cos\frac{\pi}{L},~\text{\normalfont{sign}}(\tilde{v}_{\alpha})\sqrt{r^{*2}-\cos^2\frac{\pi}{L}}\right]^\mathsf{T}, ~&\text{if }~r_0< r^*\leq 1
\end{aligned}
\right.$$
 and 
 \begin{equation}\label{opt:r}
 r^*\in\arg\min_{0\leq r\leq 1} h(r).
 \end{equation}
\end{yinli}

With the above lemma, we successfully transform the original two-dimensional problem in \eqref{eqn:sol1} into a one-dimensional problem in \eqref{opt:r}. To solve problem \eqref{opt:r}, we consider two cases $r\in[0,r_0]$ and $r\in(r_0,1]$ separately. 
If $ r\in[0, r_0]$, $h(r)$ is a quadratic function whose minimizer is given by $r_1^*=\min\{\sqrt{\tilde{u}_{\alpha}^2+\tilde{v}_{\alpha}^2}+\frac{\beta}{2}, r_0\}$. If $ r\in (r_0,1]$, $h(r)$ is convex, and thus the candidates for the minimizer are the solution to $h'(r)=0$ and the endpoints, i.e., $r_0$ and  $1$.  The solution to $h'(r)=0$ can be obtained by solving a quartic equation (which has closed-form solutions). Furthermore, we can show that the quartic equation has a unique positive solution and we denote this solution by $\bar{r}$. Then, the minimizer of $h(r)$ over $(r_0,1]$ is given by:\vspace{-0.2cm}
 $$\vspace{-0.1cm}
 r_2^*=\left\{
\begin{aligned}
\max\{\bar{r},r_0\},~&\text{if }\max\{\cos\frac{\pi}{L},\frac{\beta}{2}\}\leq \bar{r} \leq 1;\\
1,\qquad~&\text{otherwise.}
\end{aligned}\right.
$$
 The optimal solution to problem \eqref{opt:r} can then be expressed as $r^*=\arg\min_{r\in\{r_1^*,r_2^*\}} ~h(r),$ based on which we can further obtain the optimal solution to problem \eqref{eqn:sol1} using Lemma \ref{lem1}.

 Finally, from the above discussions, it is easy to check that if $\beta\geq2$, i.e., $\tau_k\leq \lambda$, then $r^*=1$ and $[u^*, v^*]$ is a feasible solution of (P).  This means that as we gradually increase the penalty parameter, we will definitely encounter a feasible solution of (P). Therefore, we can stop the AO algorithm once a feasible solution is obtained so that Algorithm 1 can terminate within finite steps and return a feasible solution.

\vspace{-0.2cm}

\section{Simulation Results}\vspace{-0.1cm}
In this section, we present numerical results to evaluate the performance of our proposed approach and the impact of quantization levels on the system performance. The entries of $\bH$ are i.i.d. following $\mathcal{C}\mathcal{N}(0,1)$ and the transmission power at the BS is $P_T=1$.
All the obtained results are averaged over $10^5$ channel realizations.
For Algorithm \ref{alg:framework}, the initial point is $\x^{(0)}=\mathbf{0}$; the initial penalty parameter and the increasing factor are $\lambda^{(0)}=\frac{0.001M}{8\sqrt{2}}$ and $\delta=5$, respectively.  For the AO algorithm, we choose $\rho_k=\rho=\frac{\sqrt{2}}{5\|\bA\|_2},~c_k=\frac{0.03}{\rho k^{0.25}},$ and $ \tau_k=\frac{6}{5\sqrt{2}}\text{mean}\left(|\bA|\right)k^{0.5}$. The AO algorithm is stopped if the distance between successive iterates is less than $10^{-2}$, or if the iteration number reaches $500$, or if a feasible solution of (P) is obtained. \vspace{-0.2cm}
\begin{figure}[t]
\centering\vspace{-0.3cm}
\subfigure{
\includegraphics[scale=0.32]{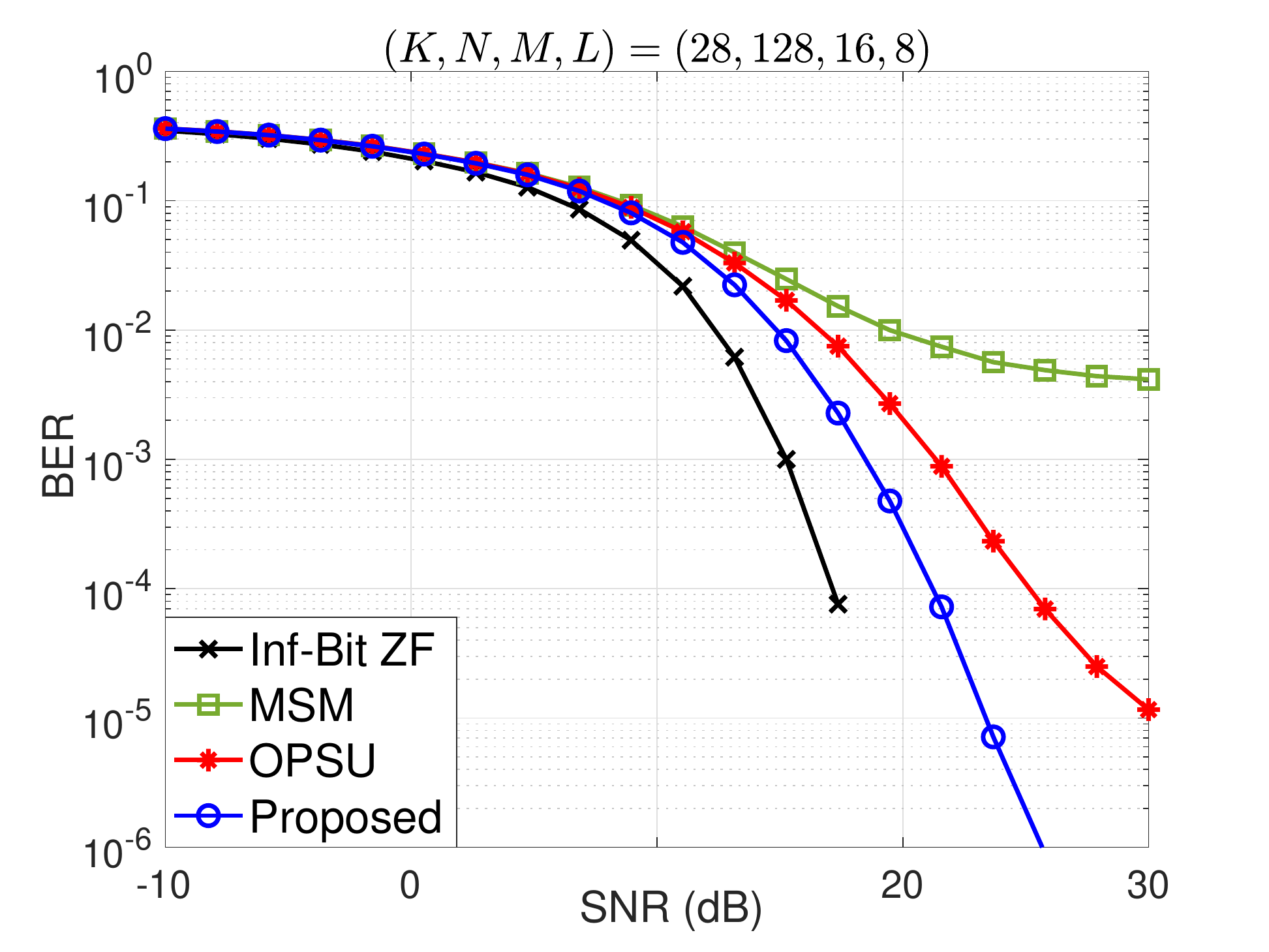}}
\vspace{-0.2cm}
\caption{BER performance of different algorithms.}
\label{ber}\vspace{-0.4cm}
\end{figure}
\begin{table}
\caption{CPU time (in sec.) of different algorithms.}
\centering
\begin{tabular}{c|ccc}
\hline
$(K,N,M,L)$&MSM&OPSU&Proposed\\
\hline
$(28,128,16,8)$&0.0904&0.0918&\textbf{0.0357}\\
\hline
\end{tabular}
\label{cputime}\vspace{-0.4cm}
\end{table}

In Fig. \ref{ber} and Table \ref{cputime}, we compare the BER performance and CPU time of our proposed algorithm with the state-of-the-art CI-based algorithms (including the MSM algorithm \cite{ciqce} and the ordered partial sequential update (OPSU) algorithm  \cite{PBB}\footnote{The OPSU algorithm is originally proposed to solve the one-bit precoding problem and here we appropriately modify it to solve the general QCE precoding problem.}). Infinite-resolution ZF precoding (where the transmit signal has only the total power constraint) is included in Fig. \ref{ber} as the performance upper bound. 
We consider a large system with $28$ users and $128$ transmit antennas,  and the number of quantization levels is set as $L=8$. As can be observed, the proposed approach exhibits significantly better BER performance with only about one-third CPU time than the MSM and OPSU algorithms, which demonstrates its superiority.

\begin{figure}[t]\vspace{-0.4cm}
\centering
\subfigure[BER versus SNR]{\includegraphics[scale=0.32]{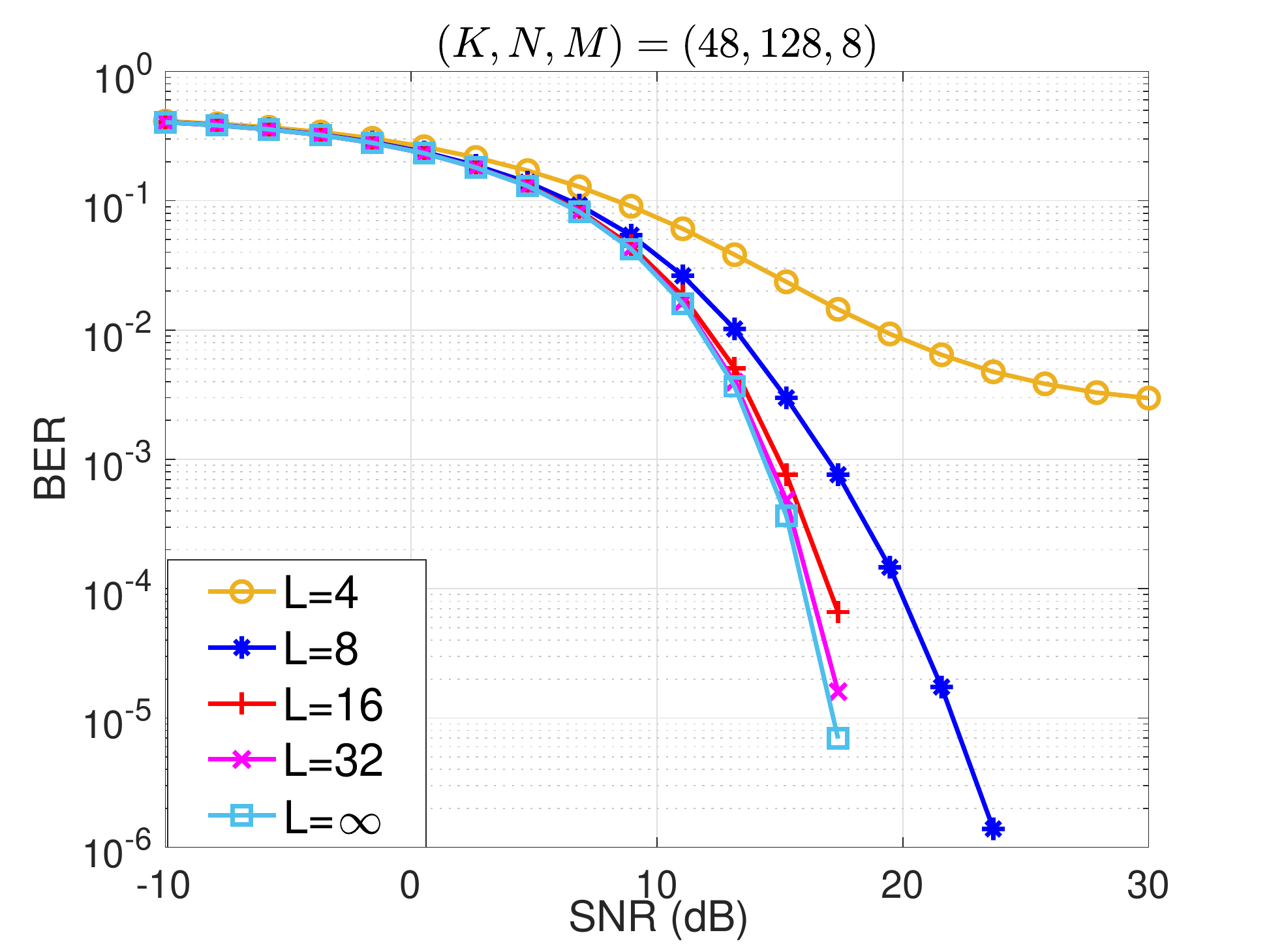}}
\subfigure[BER versus $L$]{
\includegraphics[scale=0.32]{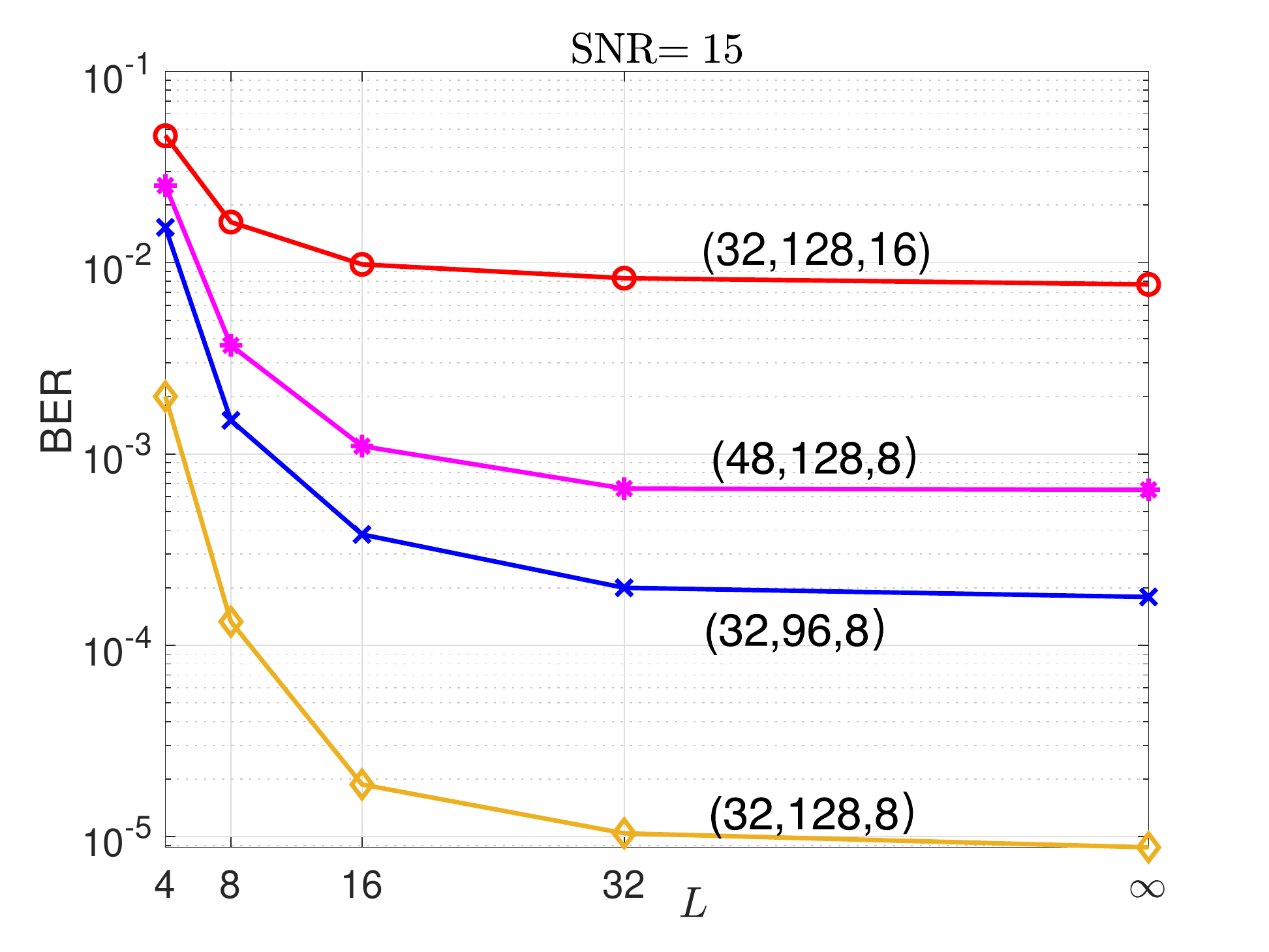}}\vspace{-0.3cm}
\caption{BER performance under different quantization levels.}
\label{bervsL}
\end{figure}\vspace{-0.0cm}
In Fig. \ref{bervsL}, we investigate the impact of the number of quantization levels (i.e., $L$)  on the system performance. Specifically, in Fig. \ref{bervsL} (a) we consider a system with $48$ users and $128$ antennas and depict the BER performance versus the SNR for different $L$s; in Fig. \ref{bervsL} (b) we fix $\text{SNR}=15$ and show the BER performance versus $L$ for different systems.  As shown in the figure, a significant performance gain can always be obtained if $L$ is increased from $4$ to $8$ (corresponding to 1-bit and 2-bits DAC quantization). There is also a slight gain if increasing $L$ from $8$ to $16$ (3-bits DAC quantization) and to $32$ (4-bits DAC quantization). However, with only $L=32$, the performance gap to the infinite-resolution case (i.e., $L=\infty$) is negligible.

From the above simulation results, we can draw the following conclusions and useful engineering insights. First, the proposed approach achieves better BER performance with less computational time than the existing algorithms in massive MIMO systems. Second, increasing the resolution of DACs from $1$ bit to $2-4$ bits in massive MIMO systems 
can generally significantly improve the system performance.



\newpage

\bibliographystyle{IEEEtran}
\bibliography{reference}

\end{document}